\documentclass[prd,twocolumn]{revtex4}
\pdfoutput=0
\usepackage{graphicx}
\usepackage{amsmath,amssymb,amsthm,amsbsy}
\usepackage{latexsym}
\usepackage{amsfonts}
\usepackage{amssymb}

\usepackage{listings}

\lstdefinestyle{mystyle}{
    basicstyle=\tiny,
    breakatwhitespace=false,         
    breaklines=true,                 
    captionpos=b,                    
    keepspaces=true,                 
    numbers=none,                    
    numbersep=5pt,                  
    showspaces=false,                
    showstringspaces=false,
    showtabs=false,                  
    tabsize=2
}
 
\newcommand{\sgn}{\mathop{\rm sgn}}

\begin{document}

\title{Quasi-Local Energy of a Rotating Object Described by Kerr Spacetime}
\author{Bjoern S. Schmekel}
\affiliation{Department of Physics, College of Studies for Foreign Diploma Recipients at the University of Hamburg, 20355 Hamburg, Germany}
\email{bss28@cornell.edu}

\begin{abstract}
The Brown-York quasi-local energy of a rotating black hole described by the Kerr metric and enclosed by a fixed-radius surface is calculated by direct
computation. No special assumptions on the angular momentum or the radial coordinate in Boyer-Lindquist coordinates were placed. 
The arbitrary reference term has been set to zero. The result may be relevant for applications in astrophysics, for modeling elementary particles
or for extensions of the framework of quasi-local quantities. An analytic expression in terms of incomplete elliptic integrals is given. 
\end{abstract}

\maketitle

\section{Introduction}
It is well understood why energy density cannot be defined locally in general relativity. Due to the equivalence principle
the influence of gravity can always be gauged away at any chosen point. However, this is only possible at a single point
and not throughout an extended region, so a meaningful expression for energy can only be defined on the quasi-local
level. A viable candidate which can be derived from an action principle is the Brown-York QLE \cite{Brown:1992br} which will be the subject
of this paper.  

So far the QLE has been computed only for few practical metrics. In this paper we extend the well-known result for the Schwarzschild metric
to the Kerr metric for spinning black holes. Since exact solutions to the Einstein field equations are in short supply \cite{ExactSpacetimes} the few
that do exist describing black holes are highly relevant for astrophysical considerations. With the recent direct observation of gravitational waves
a measure of energy and its loss may help to simplify calculations pertaining to the energy lost due to gravitational waves.
Previous results have been obtained for slowly rotating black holes in ordinary general relativity \cite{Martinez:1994ja} and in teleparallel gravity \cite{Maluf:1996kx}. 
Because of the difficulties arising from the reference term in the Brown-York QLE for certain non-embeddable spacetimes including the Kerr metric the latter
has been investigated using alternative definitions recently \cite{PhysRevD.95.084042,0264-9381-35-5-055007}. 

Black holes have been proposed as models for elementary particles \cite{Burinskii:2005mm,Lundgren:2006fu}. For a thorough understanding the values which can be assigned to mass and (self-)energy
of the particle described by a spacetime metric is crucial. 

Finally, there is still basic research being done on the framework of quasi-local quantities. Attempts are being made to find a boost-invariant
version of QLE \cite{Epp:2000zr}. Since a spinning black hole observed by an observer at rest should result in the same energy as seen by a rotating observer
watching a static Schwarzschild black hole the expressions presented in this paper may help to find suitable boost-invariant quantities.

Extensive use of computer algebra has been made in order to compute the results in this paper with most of the work being done with Maple 17 for Solaris 10 and
the add-on package GRTensor II \cite{Pollney:1996kq,GRTensorII}. Some results were double-checked with Mathematica 7 and the add-on package MathTensor 2.2.2 \cite{MathTensor}. 
A very recent upgrade to GRTensor III 2.1.11 \cite{GRTensorIII} running on Maple 18 under Linux yields the same results. 

The paper is organized as follows: First, the framework of the Brown-York quasi-local energy is reviewed. In section III issues regarding the background subtraction term are discussed
and a justification for omitting this term is given. In section IV the quasi-local quantities are computed for the Kerr metric and a fixed radius surface in Boyer-Lindquist coordinates followed by
a discussion of the results in section V.

\section{Brown-York Quasi-Local Energy}
\begin{figure}
\scalebox{0.3}{\includegraphics{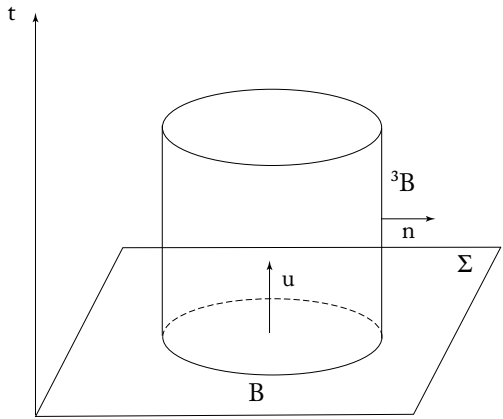}}
\caption{The time evolution of $B$ is $^3 B$. Their unit normal is $n^\mu$ in both cases. The unit normal of a time-slice
$\Sigma$ embedded in the four-dimensional manifold $M$ is denoted by $u^\mu$. Note that one dimension has been suppressed,
i.e. the boundary $B$ which is shown as a one-dimensional line represents a two-dimensional boundary.}
\label{BYsetup}
\end{figure}

A spacetime $M$ with metric $g_{\mu \nu}$, covariant derivative $\nabla_\mu$ and intrinsic curvature 
$\mathcal{R}_{\mu \nu}$ consists of timeslices $\Sigma$ whose induced metric, intrinsic and extrinsic curvature as embedded in $M$ are
denoted by  $h_{ij}$, $R_{\mu \nu}$ and $K_{\mu \nu}$, respectively. $^3 B$ is the time evolution of the
boundary $B$ as shown in Fig.~\ref{BYsetup}. The induced metric of the former is labeled $\gamma_{ij}$
and its extrinsic curvature as embedded in $M$ is denoted by $\Theta_{\mu \nu}=-\gamma_{\mu}^{\lambda} \nabla_{\lambda} n_{\nu}$. The unit normals of
$\Sigma$ and $^3 B$ are $u^\mu$ and $n^\mu$, respectively. They are assumed to satisfy the orthogonality condition $u \cdot n |_{^3 B} =0$.
We consider the energy contained in a region bounded by the two-dimensional boundary $B$.
The Brown-York surface stress-energy-momentum tensor is defined as
\begin{eqnarray}
\tau^{ij} \equiv \frac{2}{\sqrt{- \gamma}} \frac{\delta S_{cl}}{\delta \gamma_{ij}}
\label{deftau}
\end{eqnarray}
where $S_{cl}$ is the action consisting of the Einstein-Hilbert term, a potential matter term and boundary terms \cite{PhysRevLett.28.1082}
\begin{eqnarray} \nonumber
S & = & \frac{1}{2 \kappa} \int_M d^4 x \sqrt{-g} \mathcal{R} + \frac{1}{\kappa} \int_{t_i}^{t_f} d^3 x \sqrt{h} K \\ 
& - & \frac{1}{\kappa} \int_{^3 B} d^3 x \sqrt{-\gamma} \Theta - S_0[\gamma_{ij}] + S_m
\label{action}
\end{eqnarray}
evaluated at a classical solution of the Einstein field equations. This effectively suppresses the bulk term and the matter action 
and the definition of $\tau^{ij}$ is based on the presence of the boundary terms. $S_0$ is an arbitrary functional of $\gamma_{ij}$. 
Its inclusion does not alter the equations of motion and is a source of ambiguity. Finally, energy can be defined as a surface
integral
\begin{eqnarray}
E = \frac{1}{\kappa} \int_B d^2 x \sqrt{\sigma} u_i u_j \tau^{ij} = \frac{1}{\kappa} \int_B d^2 x \sqrt{\sigma} \left ( k - k_0 \right )
\label{BYenergy}
\end{eqnarray}
where $k_{\mu \nu} = -\sigma^\alpha_\mu D_\alpha n_\nu$ is the extrinsic curvature of $B$ embedded in a time-slice $\Sigma$ and the surface gravity is
denoted by $\kappa$. 
For spatial vectors the covariant derivative $D_\mu$ is defined as $D_\mu t^\nu = h^\alpha_\mu h^\nu_\beta \nabla_\alpha t^\beta$. 
The subtraction term $k_0$ depends solely on the induced metric $\sigma_{\mu \nu}$ on the boundary $B$. 
Its presence is due to the undetermined functional $S_0$. 
Alternatively,
\begin{eqnarray}
\tau^{ij} = \tau_1^{ij} + \tau_0^{ij} = -\frac{1}{\kappa} \left ( \Theta \gamma^{ij} - \Theta^{ij} \right ) - \frac{2}{\sqrt{- \gamma}} \frac{\delta S_0}{\delta \gamma_{ij}}
\end{eqnarray}
In general the index "0" refers to reference terms whereas unreferenced quantities are denoted by the index "1".
Note that $\tau^{ij}$ includes both the energy due to the gravitational field and the matter fields. 

\section{Background subtraction term}
\subsection{Choice of $S_0$}
The presence of the functional $S_0$ in the action which in subsequent equations shows up as $k_0$ and finally as a reference energy has been source of intense discussion. In the original paper by Brown and York \cite{Brown:1992br} this reference term was determined by an embedding into flat space with the latter being assigned zero energy. While this is a sensible and natural choice it
should be emphasized that it is nonetheless arbitrary.  Any choice for $S_0[\gamma_{ij}]$ including zero will reproduce the correct equations of motion. This is important because not every metric
including the Kerr metric which is used in this paper allows such an embedding \cite{Smarr1973}. Whether there is a sensible procedure which uniquely determines the background subtraction term for any given metric is
an intriguing question with progress being made recently \cite{PhysRevD.95.084042,0264-9381-35-5-055007}. Other methods may yield a unique reference term as well. For instance, the reference
term used for the Schwarzschild metric may equally be recovered from divergence cancelation arguments at large values of $r$. For more sophisticated approaches to divergence cancelation arguments by introducing counterterms to the action cf. \cite{Astefanesei:2006zd}. Counterterms may also be necessary in the study of certain metrics, e.g. the Taub-NUT family of metrics, with $u_i$ chosen to be a timelike Killing vector to ensure the
existence of proper conserved quantities \cite{Clarkson:2002uj}.

Even though the question of the subtraction term may not be finally settled this shall not prevent us from
using the QLE for real applications. In physics only energy differences are relevant quantities as long as the vacuum itself is not subject of the investigation. For any other calculation describing an observable measurement the subtraction term will drop out. 

\subsection{Conserved Quantities}
As an example consider the energy loss of a black hole spinning down. 
The surface stress-energy-momentum tensor satisfies the constraint \cite{Brown:1992br}
\begin{eqnarray}
\mathcal{D}_i \tau^{ij} = - T^{\mu \nu} n_\mu \gamma^j_\nu
\label{Divtau}
\end{eqnarray}
In order to compute the normal flux we compute the difference of the QLE contained in the considered region at two different times. 
Employing eqn. \ref{Divtau} and the product rule 
\begin{eqnarray}
\mathcal{D}_i \left ( \tau^{ij} u_j \right ) = u_j \mathcal{D}_i \tau^{ij} + \tau^{ij} \mathcal{D}_i u_j
\end{eqnarray}
where $\mathcal{D}_\mu t^\nu = \gamma^\alpha_\mu \gamma^\nu_\beta \nabla_\alpha t^\beta$ we obtain  \cite{AndyPrivate}
\begin{eqnarray} \nonumber
\int_{t_i \cap ^3 B}^{t_f \cap ^3 B} d^2 x \sqrt{\sigma} u_i u_j \tau^{ij}   = 
& - & \int_{^3 B} d^3 x \sqrt{-\gamma} \tau^{ij} \mathcal{D}_i u_j 
\\
& + & \int_{^3 B} d^3 x \sqrt{-\gamma} u_\mu n_\nu T^{\mu \nu}
\label{conslaw}
\end{eqnarray}
If $u_j$ is a Killing field with $\mathcal{D}_{(i} u_{j)}=0$ the first term on the right hand side is absent  \cite{Brown:1992br}. Even though this will not be the case in general
for the Kerr metric and our choices of normal vectors $\tau_1^{ij} \mathcal{D}_i u_j$ vanishes. Thus, assuming a $g_{\mu \nu}$ exists which describes the process of a black hole
spinning down particles have to be created ("Penrose process"). 

We interpret the difference of the QLE at two instances of time as the energy difference due to a flow of stress-energy plus
an additional contribution which we interpret as the energy loss or gain due to gravitational effects and / or changes in the background geometry in time.
 
In general a non-zero $S_0$ will lead to an additional contribution on the left hand side from $u_i u_j \tau_0^{ij}$ even for a surface with constant $r$ because of a possible change of the metric in time.
This contribution is picked up by the first term on the right hand side from a term containing $\tau_0^{ij} \mathcal{D}_i u_j$ . 

Therefore, for the remainder of this paper the background subtraction term will be omitted setting $S_0$ to zero. 

\subsection{QLE and Pseudotensors}
For yet another viewpoint on the subtraction term consider the Landau-Lifshitz pseudotensor which has been successfully used in linearized gravity to compute the energy contained in a spacelike
slice and the radiation loss. As previously discussed and suggested by their name pseudotensors are not gauge-invariant. However,
to linear order the resulting integral expressions are gauge-invariant \cite{Wald:1984}. Under the assumption $\nabla_{\mu} u_{\nu} \approx 0$ recasting the QLE 
for a metric which changes slowly in time gives

\begin{eqnarray}
 E_1
 & \approx &  \frac{1}{\kappa \Delta T}\int\limits_{^3 B} {d^3 x\sqrt { - \gamma } } u_\mu  u_\nu  \left( {\gamma ^{\mu \nu } \gamma _\sigma ^\rho  \nabla _\rho  n^\sigma   - \gamma ^{\rho \mu } \nabla _\rho  n^\nu  } \right) \nonumber \\ 
 & = & \frac{1}{\kappa \Delta T}\int\limits_{^3 B} {d^3 x\sqrt { - \gamma } } u_\mu  u_\nu  \left( {\gamma ^{\mu \nu } \gamma ^{\rho \sigma }  - \gamma ^{\nu \sigma } \gamma ^{\rho \mu } } \right)\nabla _\rho  n_\sigma  \nonumber \\    
 & = & - \frac{1}{\kappa \Delta T}\int\limits_M {d^4 x\sqrt { - g} } \nabla _\sigma  \nabla _\rho  \left[ {u_\mu  u_\nu  \left( {\gamma ^{\mu \nu } \gamma ^{\rho \sigma }  - \gamma ^{\nu \sigma } \gamma ^{\rho \mu } } \right)} \right] \nonumber \\ 
  & \approx & - \frac{1}{\kappa }\int\limits_{\Sigma}  {d^3 x\sqrt { h} } u_\mu  u_\nu \nabla _\sigma  \nabla _\rho  \left[ {  \left( {\gamma ^{\mu \nu } \gamma ^{\rho \sigma }  - \gamma ^{\nu \sigma } \gamma ^{\rho \mu } } \right)} \right] \nonumber \\ 
 \end{eqnarray}
where $\Delta T = t_f - t_i$. The structure of the last expression resembles the Landau-Lifshitz pseudotensor \cite{LandauLifshitzVol2} in vacuum which is given by
\begin{eqnarray}
2 \kappa t^{\mu \nu}_{LL} = (-g)^{-1} \left [ (-g) \left ( g^{\mu \nu} g^{\rho \sigma} - g^{\mu \rho} g^{\nu \sigma} \right ) \right ]_{,\rho \sigma} 
\end{eqnarray}
Hence, to linear order the Brown-York and the pseudotensor treatment in Cartesian Minkowski coordinates agree with the perturbed metric taking over the role of the induced metric \cite{Mars:2008tq}. 
Previous authors have pointed out that QLE and approaches based on
pseudotensors may be equivalent \cite{Chang:1998wj}. Other pseudotensors may give the same result up to a total divergence term which can be converted into a
surface integral. One example is the pseudotensor constructed from the terms of the Einstein tensor which are second order in the perturbation of flat space \cite{Wald:1984}. This residual surface
term can simply be absorbed by the subtraction term in the Brown-York QLE. Therefore, the arbitrariness of the Brown-York subtraction term is equivalent
to the arbitrary choice of a suitable pseudotensor. 

\section{Evaluation of quasi-local quantities}
We use the Kerr metric in modified Boyer-Lindquist coordinates
\begin{eqnarray}  
ds^2=- \left ( 1-\frac{2mr}{r^2+a^2 \cos^2 (\theta)} \right ) dt^2 + \nonumber \\
\frac{r^2+a^2 \cos^2 (\theta)}{r^2-2mr+a^2} dr^2 +
\left (r^2 + a^2 \cos^2 \theta \right ) d \theta^2 + \nonumber \\
\sin^2 \theta \left (   r^2 + a^2 + \frac{2mr a^2 \sin^2 \theta}{r^2 + a^2 \cos ^2 \theta}  \right ) d \phi^2 - \nonumber \\
\frac{4amr \sin^2 \theta }{r^2 + a^2 \cos^2 \theta} d \phi dt  
\end{eqnarray}
This representation of the metric is highly efficient for use with symbolic computer algebra systems \cite{Visser:2007fj} and reduces
the number of off-diagonal elements. 

As stated before the evaluation of energy and momentum densities at a single point is meaningless. Computing energy and momentum contained in a finite
region instead the results will depend on the chosen boundary. For the remainder of this paper boundaries with $r={\rm const.}$ will be used. The following unit vectors
are chosen

\begin{eqnarray}
u^\mu = \sqrt{\frac{r^2+a^2\cos^2 \theta}{r^2 -2mr +a^2 \cos^2 \theta}} \delta_t^\mu \\
n^\mu = \sqrt{\frac{r^2+a^2 -2mr}{r^2+a^2 \cos^2 \theta}} \delta_r^\mu
\end{eqnarray}
which satisfy the conditions $n_\mu n^\mu=1$, $u_\mu u^\mu=-1$ and $u_\mu n^\mu=0$. 

Evaluating with GRTensor yields

\begin{widetext}

\begin{eqnarray}
\det{\sigma} = -{\frac { \left( {a}^{2}-2\,mr+{r}^{2} \right)  \left( {a}^{4}{\chi}^{
4}+2\,{\chi}^{2}{a}^{2}{r}^{2}+{r}^{4} \right)  \left( {\chi}^{2}-1
 \right) }{{a}^{2}{\chi}^{2}-2\,mr+{r}^{2}}}

\end{eqnarray}

\begingroup
\scriptsize
\begin{eqnarray}
\epsilon & \equiv & u_i u_j \tau^{ij}=   {\frac {{a}^{4}{\chi}^{4}m-{a}^{4}{\chi}^{4}r-{a}^{4}{\chi}^{2}m-{a}^{
4}{\chi}^{2}r+5\,{a}^{2}{\chi}^{2}m{r}^{2}-3\,{a}^{2}{\chi}^{2}{r}^{3}
+3\,{a}^{2}m{r}^{2}-{a}^{2}{r}^{3}-8\,{m}^{2}{r}^{3}+8\,m{r}^{4}-2\,{r
}^{5}}{ \left( {\chi}^{2}{a}^{4}-2\,{a}^{2}{\chi}^{2}mr+{\chi}^{2}{a}^
{2}{r}^{2}+{a}^{2}{r}^{2}-2\,m{r}^{3}+{r}^{4} \right) \kappa\, \left( 
{a}^{2}{\chi}^{2}-2\,mr+{r}^{2} \right) }\sqrt {{\frac {{a}^{2}-2\,mr+
{r}^{2}}{{a}^{2}{\chi}^{2}+{r}^{2}}}}}

\label{epskerrmaple}
\end{eqnarray}
\endgroup

\begin{eqnarray} 
j_{\phi} \equiv -\sigma_{ai} u_j \tau^{ij} =  {\frac { ma \left( {a}^{2}{\chi}^{2}-{r}^{2} \right)  \left( {\chi}^{2}
-1 \right) \sqrt {{a}^{2}-2\,mr+{r}^{2}}}{\kappa\, \left( {a}^{2}{\chi
}^{2}-2\,mr+{r}^{2} \right) ^{3/2} \left( {a}^{2}{\chi}^{2}+{r}^{2}
 \right) }} \sgn \left ( 1 - \frac{2m}{r} \right )
  
\end{eqnarray}

\end{widetext}
Eqn. \ref{epskerrmaple} has been verified with MathTensor as well. Intermediate results for $\Theta^\mu_\nu$ and $\tau^{\mu \nu}$ can be found in the appendix.
Great care must be taken when simplifying expressions obtained during the course of this treatment. Because of the square root possessing branch cuts
an expression like $\sqrt{a/b}/\sqrt{a/c}$ does not simplify to $\sqrt{c/b}$ in general when $a$, $b$ and $c$ are taken from the complex plane. We try to stay
as general as possible. Only the presented analytical result for the integral in eqn. \ref{intE} is valid for real values of $a$, $m$ and $r$ only. 
Using the variable substitution $\chi=\cos \theta$ the integration over $d \theta$ giving the QLE
\begin{eqnarray}
E=2 \pi \int_0^{\pi} d \theta \sqrt{\sigma} \epsilon = 2 \pi \int_{-1}^{1} d \chi \frac{d \theta}{d \chi} \sqrt{\sigma} \epsilon
\label{intE}
\end{eqnarray}
succeeds using Maple 17. This results in a complex expression which can be expressed in terms of the incomplete elliptic integrals
\begin{eqnarray}
\mathfrak{E}(z,k) \equiv \int_0^z \frac{\sqrt{1-k^2 \zeta^2}}{\sqrt{1-\zeta^2}} d \zeta \\
\mathfrak{F}(z,k) \equiv \int_0^z \frac{1}{\sqrt{1-\zeta^2}\sqrt{1-k^2 \zeta^2}} d \zeta
\end{eqnarray}
With
\begin{eqnarray}
\Xi_E \equiv \mathfrak{E} \left( \left| a \right| \sqrt {{\frac {1}{r\, \left( 2\,m-r \right) }}} ,\sqrt {1-\frac{2m}{r}} \right) 
 \\
\Xi_F \equiv \mathfrak{F} \left( \left| a \right| \sqrt {{\frac {1}{r\, \left( 2\,m-r \right) }}}  ,\sqrt {1-\frac{2m}{r}} \right) 

\end{eqnarray}
we obtain

\begin{eqnarray}
 + \frac{{i\left| r \right|\left[ {\left( {6m - 2r} \right){{\left| r \right|}^2} - \left( {4{m^2} + {a^2}} \right)r + {a^2}m} \right]}}{{4\left| a \right|r\left( {m - \frac{r}{2}} \right)}}{\Xi _E} \nonumber \\
 + \frac{{i\left| r \right|\left[ {\left( {5m - r} \right){{\left| r \right|}^2} - \left( {6{m^2} + {a^2}} \right)r + 3{a^2}m} \right]}}{{4\left| a \right|r\left( {m - \frac{r}{2}} \right)}}{\Xi _F} \nonumber \\
 - \frac{{\left( {m - r} \right)\sqrt {{a^2} + {r^2}} \left[ {r\left( {2m - r} \right) - {a^2}} \right]}}{{4r\left( {m - \frac{r}{2}} \right)\sqrt {{a^2} - 2mr + {r^2}} }} = E
 \label{Enoref}
\end{eqnarray}

unless $\sqrt{r(2m-r)}<|a|$ and $r \le 2m$. If this condition is met the QLE diverges. Eqn \ref{Enoref} may be used to analytically continue the QLE into this undefined region
if analyticity can be imposed on the QLE. 
Nonetheless, the resulting expression is suitable for numerical evaluation. Maple input code of eqn. \ref{Enoref} can be found in the appendix. 
Direct numerical integration of eqn. \ref{intE} may also be possible, but the singular nature of the integrand makes numerical integration a challenging task. 

The unreferenced quasi-local momentum in $\phi$-direction
\begin{eqnarray}
J_{\phi} = 2 \pi \int_0^{\pi} d \theta \sqrt{\sigma} \hat  \phi^{\mu} j_{\mu}
\label{Jphinoref}
\end{eqnarray}
with
\begingroup
\scriptsize
\begin{eqnarray}
\hat \phi^{\mu} = \left [ \frac{\sin^2 \theta \left ( 2m a^2 r \sin^2 \theta + a^4 \cos^2 \theta + a^2 r^2 \cos^2 \theta +a^2 r^2 + r^4 \right ) }{r^2+a^2 \cos^2 \theta } \right ]^{-1/2} \delta_{\phi}^{\mu}
\nonumber \\
\end{eqnarray}
\endgroup
can only be evaluated numerically. Results for different angular momenta are shown in fig. \ref{Pnoref}. As suggested in \cite{Brown:1992br} we use a single topologically spherical surface surrounding the hole making 
use of the same boundary $B$ again.

\section{Results and Conclusion}

\begin{widetext}

\begin{figure}
\scalebox{0.5}{\includegraphics{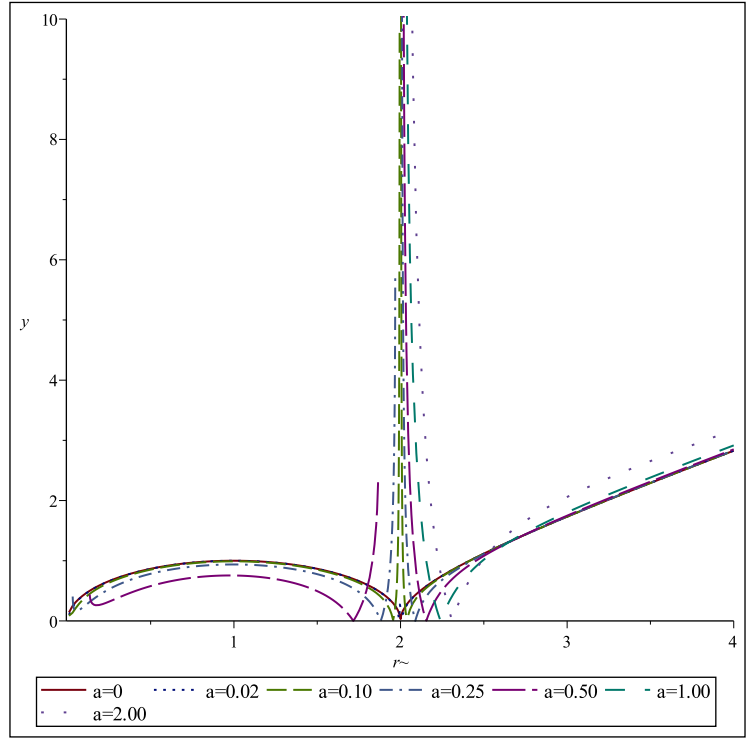}}
\caption{Absolute value of the QLE given by eqn. \ref{intE} for $m=1$ and various values of $a$}
\label{Enorefabs}
\end{figure}

\begin{figure}
\scalebox{0.5}{\includegraphics{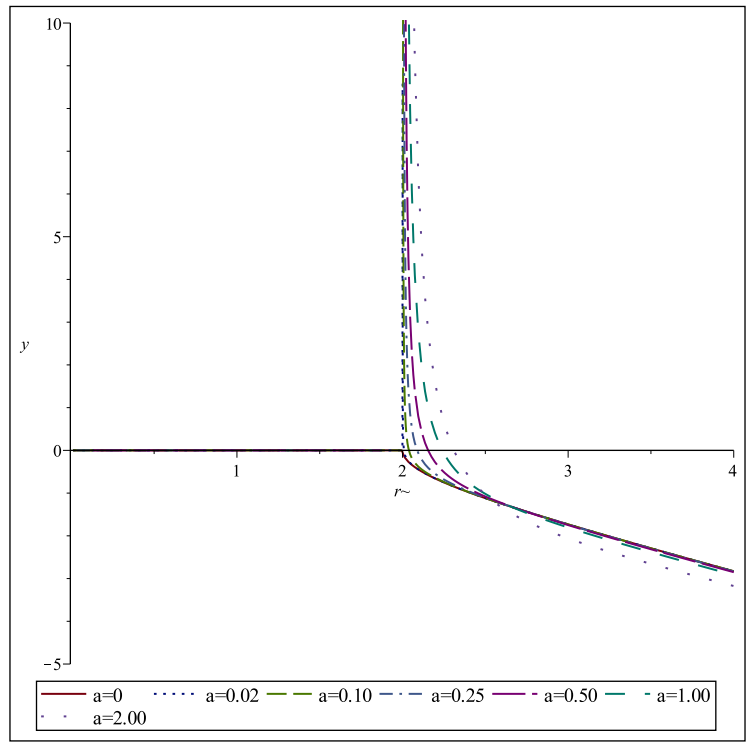}}
\caption{Real part of the QLE given by eqn. \ref{intE} for $m=1$ and various values of $a$}
\label{Enorefre}
\end{figure}

\begin{figure}
\scalebox{0.5}{\includegraphics{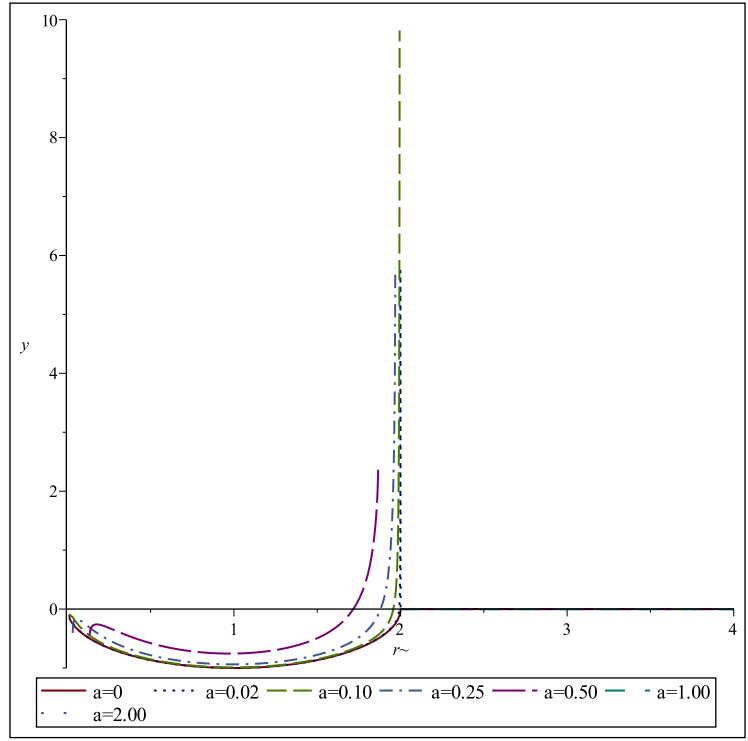}}
\caption{Imaginary part of the QLE given by eqn. \ref{intE} for $m=1$ and various values of $a$}
\label{Enorefim}
\end{figure}

\begin{figure}
\scalebox{0.5}{\includegraphics{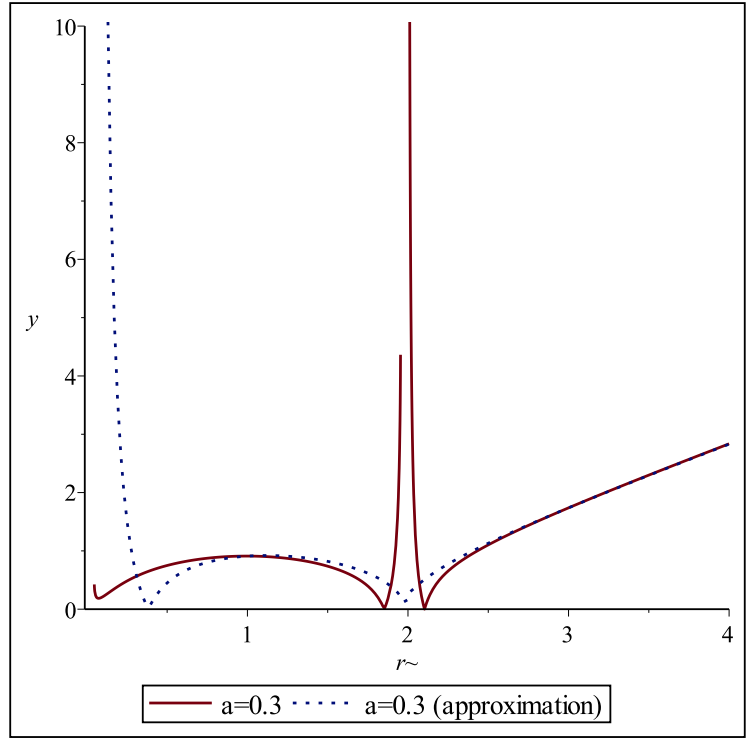}}
\caption{Comparison of the absolute value of the QLE for $m=1$ and $a=0.3$ given by eqn. \ref{intE} and the approximated result eqn. \ref{ApproxMartinez} derived by Martinez \cite{Martinez:1994ja}, respectively}
\label{Slowrot}
\end{figure}

\begin{figure}
\scalebox{0.5}{\includegraphics{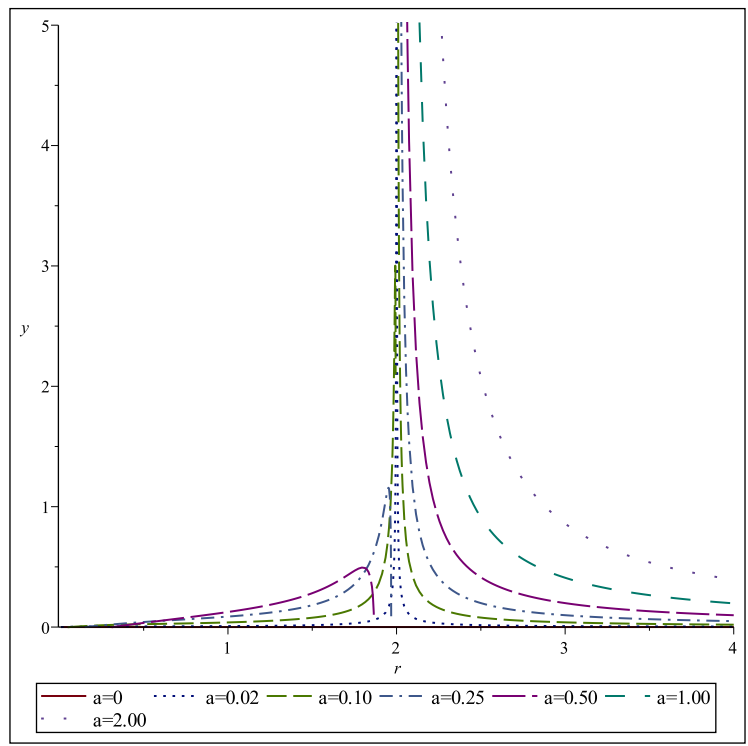}}
\caption{Quasi-local momentum in $\phi$-direction given by eqn. \ref{Jphinoref} for $m=1$ and various values of $a$}
\label{Pnoref}
\end{figure}

\end{widetext}

For $a=0$ we recover the well-known result of the unreferenced QLE for the Schwarzschild metric \cite{Brown:1992br}. With $\kappa=8\pi$ eqn. \ref{intE} yields
\begin{eqnarray}
E_1(r) = -r \sqrt{1-\frac{2m}{r}}
\end{eqnarray}
In the slow rotation regime, i.e. $|a|/m \ll 1$, the results agree with the approximation given by Martinez \cite{Martinez:1994ja}, cf. fig. \ref{Slowrot}
\begin{eqnarray}
E_1 \approx \sqrt{1-\frac{2m}{r}} \left ( 1 + \frac{a^2}{r^4 \left ( 1 - \frac{2m}{r} \right ) } \right )
\label{ApproxMartinez}
\end{eqnarray}

Since eqn. \ref{intE} may attain complex values, e.g. inside the event horizon, the absolute value as well as the real and imaginary part are plotted in fig. \ref{Enorefabs}, fig. \ref{Enorefre}  and fig. \ref{Enorefim}, respectively. 
One may feel uneasy about the energy becoming complex in certain regions wondering about the physical interpretation of this behavior. 
In classical mechanics energy is a useful quantity because it is conserved and satisfies an additivity property in the sense that the field energy of two disjoint regions
is equal to the sum of the energies contained in the respective regions. These properties still hold with eqn. \ref{conslaw} being the closest analog
to a conservation law there is. Both sides of eqn. \ref{conslaw} can become complex. In particular $\sqrt{- \gamma}$ can turn complex when spacelike
and timelike coordinates commute. It is possible to define QLE in a way that it remains real even after an interchange of timelike and spacelike coordinates \cite{Lundgren:2006fu}.
However, for the present purpose we find no advantage in doing so.

For small values of $mr^{-1}$ both eqn. \ref{intE} and the energies plotted in fig. \ref{Enorefabs} approach the value $m-r$. If one subtracted the reference term $-r$ for the Schwarzschild metric 
the ADM mass $m$ \cite{ADM:1962} would be obtained for large values of $r$ which is a consequence of the QLE measuring the gravitating mass of the black hole. In this case the reference term cancels a divergence which 
would appear otherwise as $r \longrightarrow \infty$. As pointed out earlier, though, the reference term is arbitrary and may even be omitted. For $a \neq 0$ a singularity near $r=2m$ appears in fig. \ref{Enorefabs}. 
Also, the QLE diverges very slowly for certain values of constant $m$ and $a$ as $r$ approaches zero. 

As another simple example consider the QLEs of a spinning black hole with two different angular momenta. At infinity the difference of the two energies gives zero since the QLE does not explicitly depend on $a$ in this limit as we would have expected. 

As expected the quasi-local momentum in $\phi$-direction shown in fig. \ref{Pnoref} increases with increasing $|a|$ apart from a small region in the vicinity of the "forbidden region" where $\sqrt{r(2m-r)}<|a|$ and $r \le 2m$. 
For $r=0$ and $r \longrightarrow \infty$ the momentum drops off to zero. All plotted values are real if the integral exists. 

In this paper the quasi-local energy and momentum have been computed for the Kerr metric and a surface with $r=\rm const.$ in Boyer-Lindquist coordinates with focus on the former. 
The QLE which satisfies all relevant limits could be expressed in terms of incomplete elliptic integrals. Energy differences of a spinning black hole with two different angular momenta were
considered and shown to give sensible results. The arbitrary reference term $S_0$ is not needed in this case. 

In future work the presented computations should be repeated for the Kerr-Newman metric to account for the charge of an object. Preliminary results suggest that the charge of an object may help
to attenuate the singularity of the QLE at $r=0$.

\acknowledgments
The author would like to acknowledge insightful discussion with James W. York, Jr., Andrew P. Lundgren, Shing-Tung Yau, David Brown, Stephen Lau, Lydia Bieri, Mu-Tao Wang and Dumitru Astefanesei. 
Also, he would like to mention Steven Christensen who provided excellent support for MathTensor. Research support was received from the National Science Foundation under contract \# PHY-0714648 and \# AST-0507813 during his tenure at Harvard University. 

\bibliography{bib}
\bibliographystyle{hunsrt}

\appendix
\begin{widetext}

\section{Intermediate Results}

In the listing below $\chi = \cos \theta$ has been used.

\begin{eqnarray*} 
\Theta_t^t = {\frac {m \left( {\chi}^{2}{a}^{4}+{\chi}^{2}{a}^{2}{r}^{2}-{a}^{2}{r}
^{2}-{r}^{4} \right) }{ \left( {a}^{2}{\chi}^{2}+{r}^{2} \right) 
 \left( {\chi}^{2}{a}^{4}-2\,{a}^{2}{\chi}^{2}mr+{\chi}^{2}{a}^{2}{r}^
{2}+{a}^{2}{r}^{2}-2\,m{r}^{3}+{r}^{4} \right) }\sqrt {{\frac {{a}^{2}
-2\,mr+{r}^{2}}{{a}^{2}{\chi}^{2}+{r}^{2}}}}}
 
\end{eqnarray*} 

\begin{eqnarray*}
\Theta_\theta^\theta = -{\frac {r}{{a}^{2}{\chi}^{2}+{r}^{2}}\sqrt {{\frac {{a}^{2}-2\,mr+{r}
^{2}}{{a}^{2}{\chi}^{2}+{r}^{2}}}}}
 
\end{eqnarray*} 

\begingroup
\small
\begin{eqnarray*}
\Theta_\phi^\phi = {\frac {{a}^{4}{\chi}^{4}m-{a}^{4}{\chi}^{4}r-{a}^{4}{\chi}^{2}m+{a}^{
2}{\chi}^{2}m{r}^{2}-2\,{a}^{2}{\chi}^{2}{r}^{3}+{a}^{2}m{r}^{2}+2\,m{
r}^{4}-{r}^{5}}{ \left( {a}^{2}{\chi}^{2}+{r}^{2} \right)  \left( {
\chi}^{2}{a}^{4}-2\,{a}^{2}{\chi}^{2}mr+{\chi}^{2}{a}^{2}{r}^{2}+{a}^{
2}{r}^{2}-2\,m{r}^{3}+{r}^{4} \right) }\sqrt {{\frac {{a}^{2}-2\,mr+{r
}^{2}}{{a}^{2}{\chi}^{2}+{r}^{2}}}}}
 
\end{eqnarray*} 
\endgroup

\begin{eqnarray*}
\Theta_t^\phi = {\frac { \left( {\chi}^{2}{a}^{4}-{\chi}^{2}{a}^{2}{r}^{2}-{a}^{2}{r}^
{2}-3\,{r}^{4} \right) ma \left( {\chi}^{2}-1 \right) }{ \left( {a}^{2
}{\chi}^{2}+{r}^{2} \right)  \left( {\chi}^{2}{a}^{4}-2\,{a}^{2}{\chi}
^{2}mr+{\chi}^{2}{a}^{2}{r}^{2}+{a}^{2}{r}^{2}-2\,m{r}^{3}+{r}^{4}
 \right) }\sqrt {{\frac {{a}^{2}-2\,mr+{r}^{2}}{{a}^{2}{\chi}^{2}+{r}^
{2}}}}}
 
\end{eqnarray*} 

\begin{eqnarray*}
\Theta_\phi^t = {\frac {ma \left( {a}^{2}{\chi}^{2}-{r}^{2} \right) }{ \left( {a}^{2}{
\chi}^{2}+{r}^{2} \right)  \left( {\chi}^{2}{a}^{4}-2\,{a}^{2}{\chi}^{
2}mr+{\chi}^{2}{a}^{2}{r}^{2}+{a}^{2}{r}^{2}-2\,m{r}^{3}+{r}^{4}
 \right) }\sqrt {{\frac {{a}^{2}-2\,mr+{r}^{2}}{{a}^{2}{\chi}^{2}+{r}^
{2}}}}}

\end{eqnarray*} 

\begin{eqnarray*}
\tau^{tt} = {\frac {{a}^{2}{\chi}^{2}m-{a}^{2}{\chi}^{2}r-{a}^{2}m-{a}^{2}r-2\,{r}
^{3}}{\kappa\, \left( {\chi}^{2}{a}^{4}-2\,{a}^{2}{\chi}^{2}mr+{\chi}^
{2}{a}^{2}{r}^{2}+{a}^{2}{r}^{2}-2\,m{r}^{3}+{r}^{4} \right) }\sqrt {{
\frac {{a}^{2}-2\,mr+{r}^{2}}{{a}^{2}{\chi}^{2}+{r}^{2}}}}}
 
\end{eqnarray*} 

\begin{eqnarray*}
\tau^{\theta \theta} = -{\frac {m-r}{\kappa\, \left( {\chi}^{2}{a}^{4}-2\,{a}^{2}{\chi}^{2}mr
+{\chi}^{2}{a}^{2}{r}^{2}+{a}^{2}{r}^{2}-2\,m{r}^{3}+{r}^{4} \right) }
\sqrt {{\frac {{a}^{2}-2\,mr+{r}^{2}}{{a}^{2}{\chi}^{2}+{r}^{2}}}}}
 
\end{eqnarray*} 

\begin{eqnarray*}
\tau^{\phi\phi} = -{\frac {m-r}{\kappa\, \left( {\chi}^{2}{a}^{4}-2\,{a}^{2}{\chi}^{2}mr
+{\chi}^{2}{a}^{2}{r}^{2}+{a}^{2}{r}^{2}-2\,m{r}^{3}+{r}^{4} \right) 
 \left(  1 - \chi^2  \right) }\sqrt {{\frac {{a}^{2
}-2\,mr+{r}^{2}}{{a}^{2}{\chi}^{2}+{r}^{2}}}}}
 
\end{eqnarray*} 

\begin{eqnarray*}
\tau^{t \phi} = \tau^{\phi t} = -{\frac {ma}{\kappa\, \left( {\chi}^{2}{a}^{4}-2\,{a}^{2}{\chi}^{2}mr+
{\chi}^{2}{a}^{2}{r}^{2}+{a}^{2}{r}^{2}-2\,m{r}^{3}+{r}^{4} \right) }
\sqrt {{\frac {{a}^{2}-2\,mr+{r}^{2}}{{a}^{2}{\chi}^{2}+{r}^{2}}}}}
 
\end{eqnarray*}





\section{Maple input code}
\begin{lstlisting}[language=C]
-2*Pi*(-2*piecewise((r*(2*m-r))^(1/2)/abs(a) < 1,infinity*(signum((-1/(r*(2*m-r))^(1/2))^(1/2)*signum(a^2-2*m*r+r^2)*((a^2-2*m*r+r^2)/m/r)^(1/2)*(a^2-2*m*r+r^2)^(1/2)*(r*(2*m-r))^(1/2)*m*(m-r)/(2*m-r)/kappa)-signum((r*(2*m-r))^(1/4)*signum(a^2-2*m*r+r^2)*((a^2-2*m*r+r^2)/m/r)^(1/2)*(a^2-2*m*r+r^2)^(1/2)*m*(m-r)/(2*m-r)/kappa)),0)*(a^2+r^2)^(1/2)*(a^2-2*m*r+r^2)^(1/2)*(1/r/(2*m-r))^(1/2)*r*kappa*m*abs(a)+piecewise((r*(2*m-r))^(1/2)/abs(a) < 1,infinity*(signum((-1/(r*(2*m-r))^(1/2))^(1/2)*signum(a^2-2*m*r+r^2)*((a^2-2*m*r+r^2)/m/r)^(1/2)*(a^2-2*m*r+r^2)^(1/2)*(r*(2*m-r))^(1/2)*m*(m-r)/(2*m-r)/kappa)-signum((r*(2*m-r))^(1/4)*signum(a^2-2*m*r+r^2)*((a^2-2*m*r+r^2)/m/r)^(1/2)*(a^2-2*m*r+r^2)^(1/2)*m*(m-r)/(2*m-r)/kappa)),0)*(a^2+r^2)^(1/2)*(a^2-2*m*r+r^2)^(1/2)*(1/r/(2*m-r))^(1/2)*r^2*kappa*abs(a)-2*piecewise(-1 < -(r*(2*m-r))^(1/2)/abs(a),infinity*(signum((-1/(r*(2*m-r))^(1/2))^(1/2)*signum(a^2-2*m*r+r^2)*((a^2-2*m*r+r^2)/m/r)^(1/2)*(a^2-2*m*r+r^2)^(1/2)*(r*(2*m-r))^(1/2)*m*(m-r)/(2*m-r)/kappa)-signum((r*(2*m-r))^(1/4)*signum(a^2-2*m*r+r^2)*((a^2-2*m*r+r^2)/m/r)^(1/2)*(a^2-2*m*r+r^2)^(1/2)*m*(m-r)/(2*m-r)/kappa)),0)*(a^2+r^2)^(1/2)*(a^2-2*m*r+r^2)^(1/2)*(1/r/(2*m-r))^(1/2)*r*kappa*m*abs(a)+piecewise(-1 < -(r*(2*m-r))^(1/2)/abs(a),infinity*(signum((-1/(r*(2*m-r))^(1/2))^(1/2)*signum(a^2-2*m*r+r^2)*((a^2-2*m*r+r^2)/m/r)^(1/2)*(a^2-2*m*r+r^2)^(1/2)*(r*(2*m-r))^(1/2)*m*(m-r)/(2*m-r)/kappa)-signum((r*(2*m-r))^(1/4)*signum(a^2-2*m*r+r^2)*((a^2-2*m*r+r^2)/m/r)^(1/2)*(a^2-2*m*r+r^2)^(1/2)*m*(m-r)/(2*m-r)/kappa)),0)*(a^2+r^2)^(1/2)*(a^2-2*m*r+r^2)^(1/2)*(1/r/(2*m-r))^(1/2)*r^2*kappa*abs(a)+6*EllipticF((1/r/(2*m-r))^(1/2)*abs(a),(-1/r*(2*m-r))^(1/2))*a^2*m*(-(a^2-2*m*r+r^2)/r/(2*m-r))^(1/2)*(a^2+r^2)^(1/2)*abs(r)-2*EllipticF((1/r/(2*m-r))^(1/2)*abs(a),(-1/r*(2*m-r))^(1/2))*a^2*r*(-(a^2-2*m*r+r^2)/r/(2*m-r))^(1/2)*(a^2+r^2)^(1/2)*abs(r)-12*EllipticF((1/r/(2*m-r))^(1/2)*abs(a),(-1/r*(2*m-r))^(1/2))*m^2*r*(-(a^2-2*m*r+r^2)/r/(2*m-r))^(1/2)*(a^2+r^2)^(1/2)*abs(r)+10*m*(-(a^2-2*m*r+r^2)/r/(2*m-r))^(1/2)*(a^2+r^2)^(1/2)*abs(r)^3*EllipticF((1/r/(2*m-r))^(1/2)*abs(a),(-1/r*(2*m-r))^(1/2))-2*(-(a^2-2*m*r+r^2)/r/(2*m-r))^(1/2)*(a^2+r^2)^(1/2)*abs(r)^3*r*EllipticF((1/r/(2*m-r))^(1/2)*abs(a),(-1/r*(2*m-r))^(1/2))+2*EllipticE((1/r/(2*m-r))^(1/2)*abs(a),(-1/r*(2*m-r))^(1/2))*a^2*m*(-(a^2-2*m*r+r^2)/r/(2*m-r))^(1/2)*(a^2+r^2)^(1/2)*abs(r)-2*EllipticE((1/r/(2*m-r))^(1/2)*abs(a),(-1/r*(2*m-r))^(1/2))*a^2*r*(-(a^2-2*m*r+r^2)/r/(2*m-r))^(1/2)*(a^2+r^2)^(1/2)*abs(r)-8*EllipticE((1/r/(2*m-r))^(1/2)*abs(a),(-1/r*(2*m-r))^(1/2))*m^2*r*(-(a^2-2*m*r+r^2)/r/(2*m-r))^(1/2)*(a^2+r^2)^(1/2)*abs(r)+12*m*(-(a^2-2*m*r+r^2)/r/(2*m-r))^(1/2)*(a^2+r^2)^(1/2)*abs(r)^3*EllipticE((1/r/(2*m-r))^(1/2)*abs(a),(-1/r*(2*m-r))^(1/2))-4*(-(a^2-2*m*r+r^2)/r/(2*m-r))^(1/2)*(a^2+r^2)^(1/2)*abs(r)^3*r*EllipticE((1/r/(2*m-r))^(1/2)*abs(a),(-1/r*(2*m-r))^(1/2))-2*m*(1/r/(2*m-r))^(1/2)*abs(a)^5+2*r*(1/r/(2*m-r))^(1/2)*abs(a)^5+4*r*m^2*(1/r/(2*m-r))^(1/2)*abs(a)^3-8*m*r^2*(1/r/(2*m-r))^(1/2)*abs(a)^3+4*r^3*(1/r/(2*m-r))^(1/2)*abs(a)^3+4*m^2*r^3*(1/r/(2*m-r))^(1/2)*abs(a)-6*m*r^4*(1/r/(2*m-r))^(1/2)*abs(a)+2*r^5*(1/r/(2*m-r))^(1/2)*abs(a))/(a^2+r^2)^(1/2)/(a^2-2*m*r+r^2)^(1/2)/(1/r/(2*m-r))^(1/2)/r/(2*m-r)/kappa/abs(a)
\end{lstlisting}

\end{widetext}

\end{document}